\documentclass[journal]{IEEEtran}
\usepackage{graphicx}
\usepackage{amssymb}
\usepackage{amsfonts}
\usepackage{amsmath}
\usepackage{epsfig}
\usepackage{color}
\usepackage{fancybox}
\usepackage{textcomp}
\usepackage{multirow}
\usepackage{setspa ce}
\usepackage{psfrag}
\usepackage{booktabs}
\usepackage{float}
\usepackage{array}

\usepackage{color}

\begin{document}
	%

	\title{Deep Learning-based CSI Feedback Approach for Time-varying Massive MIMO Channels}
	%
	%
	%
	\author{Tianqi~Wang,~Chao-Kai~Wen,~Shi~Jin, Geoffrey Ye Li
				\thanks{T. Wang, and S. Jin are with the National
					Mobile Communications Research Laboratory, Southeast University, Nanjing
					210096, China (e-mail: wangtianqi@seu.edu.cn; hqwanglyt@seu.edu.cn; jinshi@seu.edu.cn).}
				\thanks{C.-K. Wen is with the Institute of Communications Engineering, National
					Sun Yat-sen University, Kaohsiung 80424, Taiwan (e-mail: ckwen@ieee.org).}
				\thanks{G.~Y.~Li is with the School of Electrical and Computer Engineering,
					Georgia Institute of Technology, Atlanta, GA 30332 USA (e-mail:
					liye@ece.gatech.edu).}
	}	
	\vspace{-3cm}
	\maketitle
	\begin{abstract}
		Massive multiple-input multiple-output (MIMO) systems rely on channel state information (CSI) feedback to perform precoding and achieve performance gain in frequency division duplex (FDD) networks. However, the huge number of antennas poses a challenge to conventional CSI feedback reduction methods and leads to excessive feedback overhead. In this article, we develop a real-time CSI feedback architecture, called CsiNet-long short-term memory (LSTM), by extending a novel deep learning (DL)-based CSI sensing and recovery network. CsiNet-LSTM considerably enhances recovery quality and improves trade-off between compression ratio (CR) and complexity by directly learning spatial structures combined with time correlation from training samples of time-varying massive MIMO channels. Simulation results demonstrate that CsiNet-LSTM outperforms existing compressive sensing-based and DL-based methods and is remarkably robust to CR reduction.
	\end{abstract}
	\begin{IEEEkeywords}
		Massive MIMO, FDD, CSI feedback, compressive sensing, deep learning.		
	\end{IEEEkeywords}		
	\setlength{\parskip}{-1em}
	%
	\setlength{\parskip}{0em}
	\IEEEpeerreviewmaketitle
	\vspace{-0.3cm}
	\section{Introduction}
	Massive multiple-input multiple-output (MIMO) systems have been recognized as a critical development for future wireless communications. With downlink channel state information (CSI), a base station (BS) with massive antennas can use channel-adaptive techniques to eliminate inter-user interference and increase channel capacity. In frequency division duplex (FDD) networks, downlink CSI can only be estimated at user equipment (UE) and fed back to the BS. The excessive overhead has motivated many feedback reduction techniques, such as vector quantization and codebook-based approaches  \cite{Love2008An}. However, quantization errors pose a challenge to CSI-sensitive applications, whereas the huge number of antennas complicates the codebook design and accordingly increases feedback overhead.
	
	The compressive sensing (CS)-based CSI feedback approaches proposed recently address the aforementioned problems by using the spatial and temporal correlation of CSI. These methods sparsify CSI under certain bases to apply CS for feedback and reconstruction \cite{Kuo2012Compressive} or distributed compressive channel estimation \cite{Rao2014Distributed}. In reality, CSI is only approximately sparse under elaborate base selection or sparsity modeling. Many existing CS algorithms experience difficulty in CSI compression and recovery if there is a model mismatch.
	
	Time correlation property of slow-varying channels has been considered in \cite{Kuo2012Compressive} to further reduce feedback quantity. This method reuses the previously retained channel information for subsequent CSI recovery if the error is under a certain threshold. However, the reused information only provides an estimate and is hard to update in real time. As a result, resolution degrades and the feedback overhead cannot reduce any more in fast-changing channels.
	
	Recently, deep learning (DL) methods has been successfully applied in wireless communications \cite{Wang2017Deep,8052521,He2018Deep}. A CSI feedback network, called CsiNet \cite{Wen2017Deep}, uses an autoencoder-like architecture to mimic the CS and reconstruction processes. It uses an encoder to obtain compressed representation (codewords) by directly learning channel structures from the training data and a decoder to recover CSI via one-off feedforward multiplication. CsiNet remarkably outperforms the CS-based methods. But it ignores time correlation in time-varying channels, and reconstruct CSI independently.
	
	In this article, we propose an improved architecture by considering time correlation. Our work is motivated by the recurrent convolutional neural network (RCNN) architecture that has been successfully used in video representation and reconstruction \cite{Xu2018CSVideoNet}. The basic idea is to use a convolutional neural network (CNN) and a recurrent neural network (RNN) to extract spatial features and interframe correlation, respectively. Our contribution in this article is summarized as follows.
	\begin{itemize}
		\item{} We propose an DL-based CSI feedback protocol for FDD MIMO systems by extending CsiNet with a long short-term memory (LSTM) network, which is a classic type of RNN. The proposed network, called CsiNet-LSTM, modifies the CNN-based CsiNet for CSI compression and initial recovery and uses LSTM to extract time correlation for further improvement in resolution.
		\item{} The experiment results demonstrate that CsiNet-LSTM achieves the best recovery quality and outperforms state-of-the-art CS methods in terms of complexity. CsiNet-LSTM exhibits remarkable robustness to compression ratio (CR) reduction and enables real-time and extensible CSI feedback applications without considerably increasing overhead compared with CsiNet.
	\end{itemize}

	%
	%

	
	\vspace{-0.3cm}
	\section{System model}
	\label{system_model}
	An FDD downlink massive MIMO-orthogonal frequency division multiplexing (OFDM) system with $N_{c}$ subcarriers is considered. The BS deploys $N_{t}$ transmit antennas as uniform linear array (ULA). In a time-varying channel caused by UE mobility, the received signal at time $t$ on the $n$th subcarrier for UE with a single receiver antenna can be modeled as,	
	\vspace{-0.1cm}
	\begin{equation}
		\begin{aligned}
			{y_{n,t}}={{\bf h}_{n,t}^{T}{\bf v}_{n,t}x_{n,t}+z_{n,t}},
			\label{mimo-ofdm}
		\end{aligned}
	\end{equation}
	where ${\bf h}_{n,t}\in \mathbb{C}^{N_{t}\times 1}$, $x_{n,t}\in \mathbb{C}$, and $z_{n,t}\in \mathbb{C}$ denote the instantaneous channel vector in the frequency domain, transmit data symbol, and additive noise, respectively, ${\bf v}_{n, t} \in \mathbb{C}^{N_{t}\times 1}$ is the beamforming or precoding vector designed by the BS based on the received downlink CSI. We denote the CSI matrix at time $t$ in the spatial-frequency domain as ${\bf H}_{t} = [{\bf h}_{1,t}, \ldots, {\bf h}_{N_{c},t}]^{T} \in \mathbb{C}^{N_{c} \times N_{t}}$. In practice, the UE continuously estimates and feeds instantaneous CSI (i.e., ${\bf H}_{t}, {\bf H}_{t+1}, \ldots$) back to the BS to track the time-varying characteristics of the channel. To reduce feedback overhead, we can exploit the following observations.
	
	\setlength{\parskip}{0.5em}
	{\bf Observation 1} (angular-delay domain sparsity):
	${\bf H}_{t}$ can be transformed into an approximately sparsified matrix ${\bf  H}'_{t}$ in the angular-delay domain via 2D discrete Fourier transform (2D-DFT) \cite{Wen2017Deep} by ${\bf  H}'_{t}={\bf F}_{\sf d}{\bf H}_{t}{\bf F}_{\sf a}$, where ${\bf F}_{\sf d} \in \mathbb{C}^{N_{c} \times N_{c}}$ and ${\bf F}_{\sf a} \in \mathbb{C}^{N_{t} \times N_{t}}$ are two DFT matrices. First, due to limited multipath time delay, performing DFT on frequency domain channel vectors (i.e., column vectors of ${\bf H}_{t}$) can transform ${\bf H}_{t}$ into a sparsed matrix in the delay domain, with only the first $N_{c}'$ ($<N_{c}$) rows having distinct non-zero values. Secondly, as proved in \cite{Wen2015Channel}, the channel matrix is sparse in a defined angle domain by performing DFT on spatial domain channel vectors (i.e., row vectors of ${\bf H}_{t}$) if the number of transmit antennas, $N_{t} \to +\infty $, is very large. Usually,  ${\bf H}_{t}'$ is only approximately sparse for finite $N_{t}$, which challenges conventional CS methods. Therefore, we will propose a DL-based feedback architecture without sparsity prior constraint. We perform sparsity transformation to decrease parameter overhead and training complexity. We retain the first $N_{c}'$ non-zero rows and truncate ${\bf H}_{t}'$ to a $N_{c}' \times N_{t}$ matrix, ${\bf H}_{t}''$, which reduces the total number of parameters for feedback to $N=N_{c}'N_{t}$.
	\setlength{\parskip}{0.3em} 	
	
	{\bf Observation 2} (correlation within coherence time):
	UE motion during communication results in a Doppler spread, i.e., time-varying characteristics of wireless channels.
	With the maximum movement velocity denoted as $v$, coherence time can be calculated as
	\vspace{-0.2cm}
	\begin{equation}
		\begin{aligned}
			\Delta t=\frac{c}{2vf_{0}},
			\label{coherence time}
		\end{aligned}
	\end{equation}
	where $f_{0}$ is the carrier frequency, and $c$ is the velocity of light. The CSI within $\Delta t$ is considered correlated with one other. Therefore, instead of independently recovering CSI, the BS can combine the feedback and previous channel information for the subsequent reconstruction. We set the feedback time interval as $\delta t$ and place $T$ adjacent instantaneous angular-delay domain channel matrices into a channel group, i.e., $\{{\bf H}_{t}''\}_{t=1}^{T}=\{{\bf H}_{1}'', \ldots, {\bf H}_{t}'', \ldots, {\bf H}_{T}''\}$. The group exhibits correlation property, as long as $T$ satisfies $ 0\leq \delta t \cdot T \leq \Delta t $.
	

	In this article, we design an encoder, ${\bf s}_{t}=f_{\sf en}({\bf H}_{t}'')$, at the UE to compress each complex-valued ${\bf H}_{t}''$ of $\{{\bf H}_{t}''\}_{t=1}^{T}$ into an $M$-dimensional real-valued codeword vector ${\bf s}_{t}$ ($M<N$). If two real number matrices are used to represent the real and imaginary parts of ${\bf H}_{t}''$, then CR will be $M/2N$. We also design a decoder with a memory that can extract time correlation from the previously recovered channel matirces, ${\hat{\bf H}}_{1}'',...,{\hat{\bf H}}_{t-1}''$,  and combine them with the received ${\bf s}_{t}$ for current reconstruction, ${\hat{\bf H}}_{t}''=f_{\sf de}({\bf s}_{t}; {\hat{\bf H}}_{1}'',...,{\hat{\bf H}}_{t-1}'')$, where $1 \leq t \leq T.$
	Then, inverse 2D-DFT is performed to obtain the original spatial frequency channel matrix.
	
	\begin{figure*}
		\centering
		\includegraphics[width=1.5in,height=2.34in]{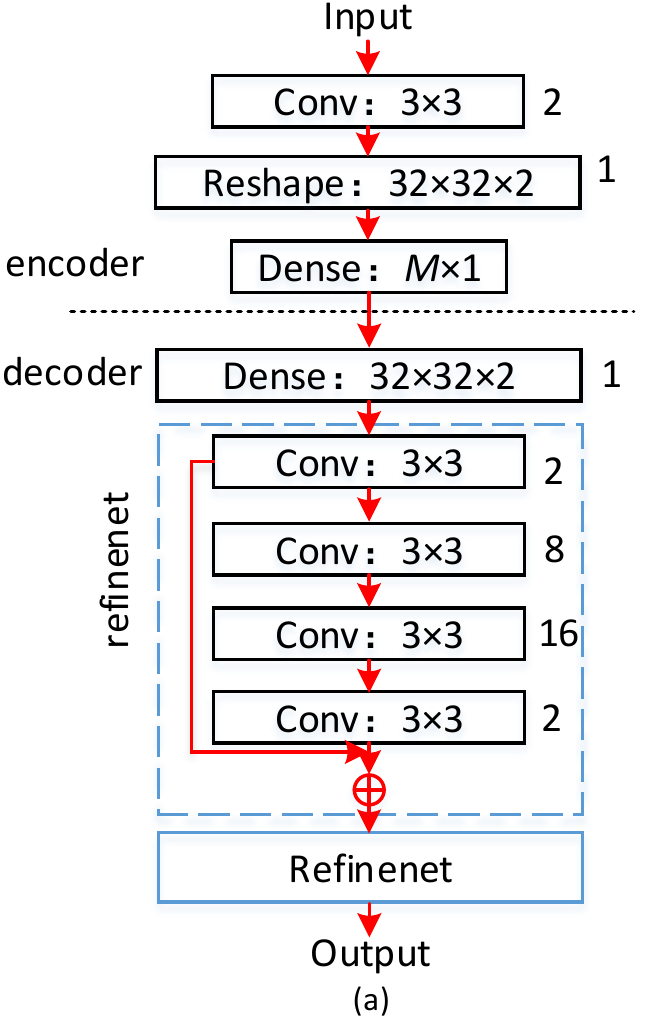}\includegraphics[width=3.6in,height=2.34in]{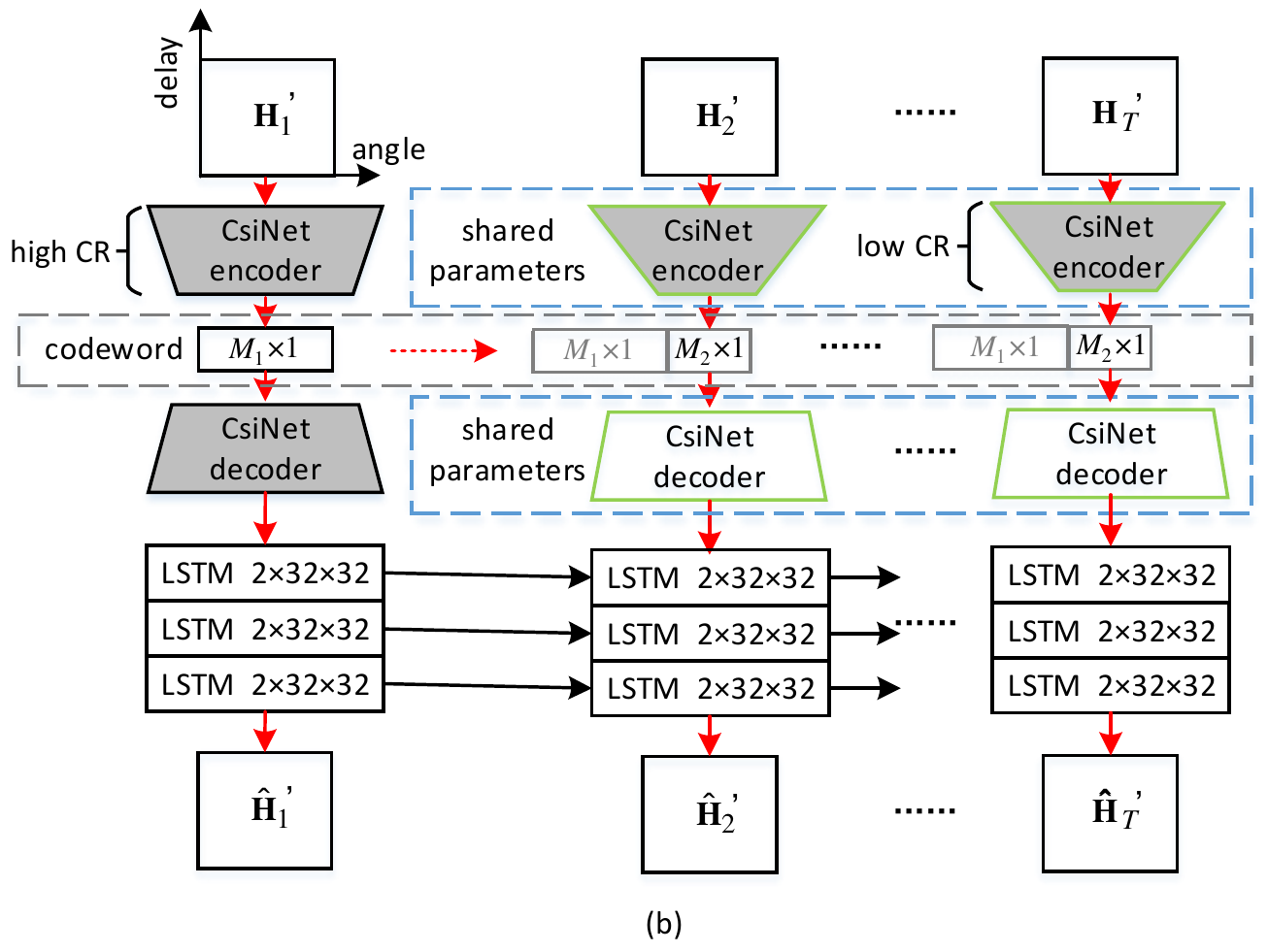}
		\vspace{-0.3cm}
		\caption{(a) CsiNet architecture presented in \cite{Wen2017Deep}. It comprises an encoder with a $ 3 \times 3 $ conv layer and an $M$-unit dense layer for sensing and a decoder with a $ 2N_{c}'N_{t} $-unit dense layer and two RefineNet for reconstruction.  Each RefineNet contains four $ 3 \times 3 $ conv layers with different channel sizes. (b) Overall architecture of CsiNet-LSTM. ${\bf H}_{1}'$ and the remaining ${T-1}$ channel matrices are compressed by high-CR and low-CR CsiNet encoders, respectively. Codewords are concatenated before being fed into the low-CR CsiNet decoder, and final reconstruction is performed by three $ 2N_{c}'N_{t} $-unit LSTMs. }		
				\vspace{-0.5cm}
		\label{system architecture}	
	\end{figure*}
\vspace{-0.3cm}
	\section{CsiNet-LSTM}
	\label{csinet-lstm}
	\setlength{\parskip}{0.1em}
	The CsiNet in \cite{Wen2017Deep} demonstrates remarkable performance in CSI sensing and reconstruction. However, the resolution degrades at low CR because the it only focuses on angular-delay domain sparsity (Observation 1) and ignores the time correlation (Observation 2) of time-varying massive MIMO channels. The two observations in Section \ref{system_model} are similar to the spatial structure and interframe correlation of videos, respectively. Motivated by RCNN that excels in extracting spatial-temporal features for video representation \cite{Xu2018CSVideoNet}, we will extend CsiNet with LSTM to improve CR and recovery quality trade-off. We will also introduce the multi-CR strategy in \cite{Xu2018CSVideoNet} to implement variable CRs on different channel matrices.
	
	The proposed CsiNet-LSTM is illustrated in Fig. \ref{system architecture}(b), with CsiNet shown in Fig. \ref{system architecture}(a).
	Our model includes the following two steps: angular-delay domain feature extraction and correlation representation and final reconstruction.
	\subsubsection{Angular-delay domain feature extraction}
	We apply CsiNet with two different CRs to $\{{\bf H}_{t}''\}_{t=1}^{T}$ to learn the angular-delay domain structure and perform sensing and initial reconstruction. A high-CR CsiNet transforms the first channel ${\bf H}_{1}''$ into an ${M_{1} \times 1}$ codeword vector that retains sufficient structure information for high resolution recovery. A low-CR CsiNet encoder performs on the remaining ${T-1}$ channel matrices to generate a series of ${M_{2} \times 1}$ codewords ($M_{1}>M_{2}$), given that less information is required due to channel correlation.
	The ${T-1}$ codewords are all concatenated with the first $M_{1}\times 1$ codeword before being fed into the low-CR CsiNet decoder to fully utilize feedback information. Each CsiNet outputs two matrices with size $N_{c}' \times N_{t}$ as extracted features from the angular-delay domain.
	
	All low-CR CsiNets shown in Fig. \ref{system architecture}(b) share the same network parameters, i.e., weights and bias, because they perform the same work. This condition dramatically reduces parameter overhead and if the value of $T$ changes to adapt to the channel-changing speed and feedback frequency, the architecture can be easily rescaled to perform on channel groups with different $T$. In practice, a low-CR CsiNet will be reused ${T-1}$ times instead of making ${T-1}$ copies. The grey blocks in Fig. \ref{system architecture}(b) load parameters from the original CsiNets as pretraining before end-to-end training with the entire architecture. This method can alleviate vanishing gradient problems due to long paths from CsiNets to LSTMs.

	\subsubsection{Correlation representation and final reconstruction}
	We use LSTMs to extend the CsiNet decoders for time correlation extraction and final reconstruction. LSTMs have inherent memory cells and can keep the previously extracted information for a long period for later prediction. In particular, the outputs of the CsiNet decoders form length $T$ sequences before being fed into three-layer LSTMs. Each LSTM has $2N_{c}'N_{t}$ hidden units, which is the same as the output dimension. The final outputs are then reshaped into two $N_{c}' \times N_{t}$ matrices as the final recovered ${\hat {\bf H}_{t}''}$. The spatial frequency domain CSI can then be obtained via inverse 2D-DFT. At each time step, the LSTMs implicitly learn time correlation from the previous inputs and then merge them with the current inputs to increase low CR recovery quality. Correlation information is updated after each step due to the nature of LSTM. The experimental results show that the highly compressed ${T-1}$ matrices can achieve better recovery accuracy than ${\bf H}_{1}''$ as a benefit from LSTMs.
	
	We use end-to-end learning to obtain all parameters for the encoder and the decoder denoted as $\Theta=\{\Theta_{\sf en},\Theta_{\sf de}\}$. Notably, ${\bf H}_{t}''$ are normalized with all elements scaled into the $[0,1]$ range before being fed into the network. This normalization is required for CsiNet. For details, we refer to \cite{Wen2017Deep}.
    Let $f$ denote the final trained network defined as
	\begin{equation*}
			\hat{\bf H}_{t}'' =
			f({\bf H}_{t}'';\Theta)		
			=f_{\sf de}(f_{\sf en}({\bf H}_{1}'';\Theta_{\sf en}), \ldots, f_{\sf en}({\bf H}_{t}'';\Theta_{\sf en});\Theta_{\sf de}).	
			\label{f}
	\end{equation*}
We select ADAM as the optimization algorithm and use mean-squared error (MSE) as the loss function, which is defined as,
	\vspace{-0.2cm}
	\begin{equation}
		\begin{aligned}
			L(\Theta)= \frac{1}{M}\sum_{m=1}^M\sum_{t=1}^T\|f({\bf H}_{t}'';\Theta)-{\bf H}_{t}''\|_{2}^{2},  			
			\label{loss function}
		\end{aligned}
	\end{equation}	
	where $M$ is the total number of samples in the training set and $\| \cdot \|_{2}$ is the Euclidean norm.

	The procedure for CsiNet-LSTM is described as follows. Multiple CR CsiNet encoders are deployed at each UE, whereas the CsiNet decoders and LSTMs are deployed at the BS. Each side has a counter. At the beginning, ${\bf H}_{1}''$ is compressed with high CR at the UE and recovered by a high-CR CsiNet decoder and initialized by the LSTMs at the BS. In the subsequent time step $t$ ($2 \leq t \leq T$), ${\bf H}_{t}''$ is transformed into a lower-dimensional codeword ${\bf s}_{t}$ at the UE, which is expected to contain the learned correlation information. The lower-dimensional codeword, ${\bf s}_{t}$ is then concatenated with the first one ${\bf s}_{1}$ and inversely transformed by the LSTMs at the BS. After each time step, the counters add by one. Similar operations continue until the counters accumulate to $T$, and the LSTMs are reset for the subsequent channel group recovery.
		
	\vspace{-0.3cm}
	\section{Simulation results and analysis}
	
	\label{experiment}
	We use the COST 2100 model \cite{Liu2012The} to simulate time-varying MIMO channels and generate training samples. We set the MIMO-OFDM system to work on a 20 MHz bandwidth with $N_{c}=256$ subcarriers and use ULA with $N_{t}=32$ antennas at the BS. The angular-delay domain channel matrix is truncated to a size of $32 \times 32$. Two scenarios are considered: the indoor scenario at 5.3 GHz with UE velocity $v=0.0036$ km/h and the outdoor scenario at 300 MHz with UE velocity $v=3.24$ km/h. Therefore, $ \Delta t$ is 30s and 0.56s, respectively. Compressed CSI is fed back every $ \delta t = 0.04$ s. We set the channel group size $T=10$, which satisfies $\delta t \cdot T < \Delta t$ in both scenarios. We perform experiments at CR values of 1/16, 1/32, and 1/64, with the first channel ${\bf H}_{1}''$ compressed under 1/4.
	
	Training, validation, and testing sets have 75,000, 12,500, 12,500 samples, respectively. Some parameters are preloaded from the CsiNet for initialization. The epochs are adjusted for a convergence situation ranging from 500 to 1,000. The batch size is 100 and the learning rates are 0.001 and 0.0001 for the former and latter epochs, respectively.
	
	We compare our architecture with three state-of-the-art CS-based algorithms, namely,  LASSO $\ell_{1}-$solver \cite{Daubechies}, TVAL3 \cite{Li2009User}, and BM3D-AMP \cite{Metzler2016From}, and the DL-based CsiNet \cite{Wen2017Deep}. LASSO uses simple sparsity priors but achieves good performance. TVAL3 is a minimum total variation method that provides remarkable recovery quality but with high computing efficiency. BM3D-AMP achieves the most accurate recovery performance on natural images and runs 10 times faster than other iterative methods.
	
	We use the default configuration in the open source codes of the aforementioned methods for simulation. When comparing with CsiNet, we consider the slight difference between datasets and refine the CsiNet parameters on our training set for several epochs for fairness. We run the conventional CS-based methods on an Intel$\circledR$Core$^{\mathrm{TM}}$ i7-6700 CPU due to the lack of a GPU solution. CsiNet and CsiNet-LSTM are trained and tested on Nvidia GeForce GTX 1080 Ti GPU.

	Normalized MSE (NMSE) is used to evaluate the recovery performance, which is defined as follows:
	\vspace{-0.1cm}
	\begin{equation}
		\begin{aligned}
			{\rm NMSE}=\mathbb{E}\bigg \{\frac{1}{T}\sum_{t=1}^{T} {\|{\bf H}_{t}''-\hat {\bf H}_{t}''\|_{2}^{2}}/{\|{\bf H}_{t}''\|_{2}^{2}}\bigg \}.
			\label{nmse}
		\end{aligned}
	\end{equation}	

	To compare with CsiNet, the following cosine similarity is also calculated:
	\vspace{-0.3cm}
	\begin{equation}
		\begin{aligned}
			\rho = \mathbb{E}\Bigg \{\frac{1}{T}\frac{1}{N_{c}} \sum_{t=1}^{T}\sum_{n=1}^{N_{c}}\frac{|\hat {\bf h}_{n,t}^{H}{\bf h}_{n,t}|}{\|\hat {\bf h}_{n,t}\|_{2}\|{\bf h}_{n,t}\|_{2}} \Bigg \},
			\label{corr}
		\end{aligned}
	\end{equation}
	where $\hat {\bf h}_{n,t}$ denotes the reconstructed channel vector of the $n$th subcarrier at time $t$. When the BS uses ${\bf v}_{n,t}={\hat {\bf h}_{n,t}/\|\hat {\bf h}_{n,t}\|_{2}}$ as a beamforming vector, $\rho$ can be used to indicate the beamforming gain.
	
\begin{figure}
			\centering
			\includegraphics[width=3.4in,height=1.8in]{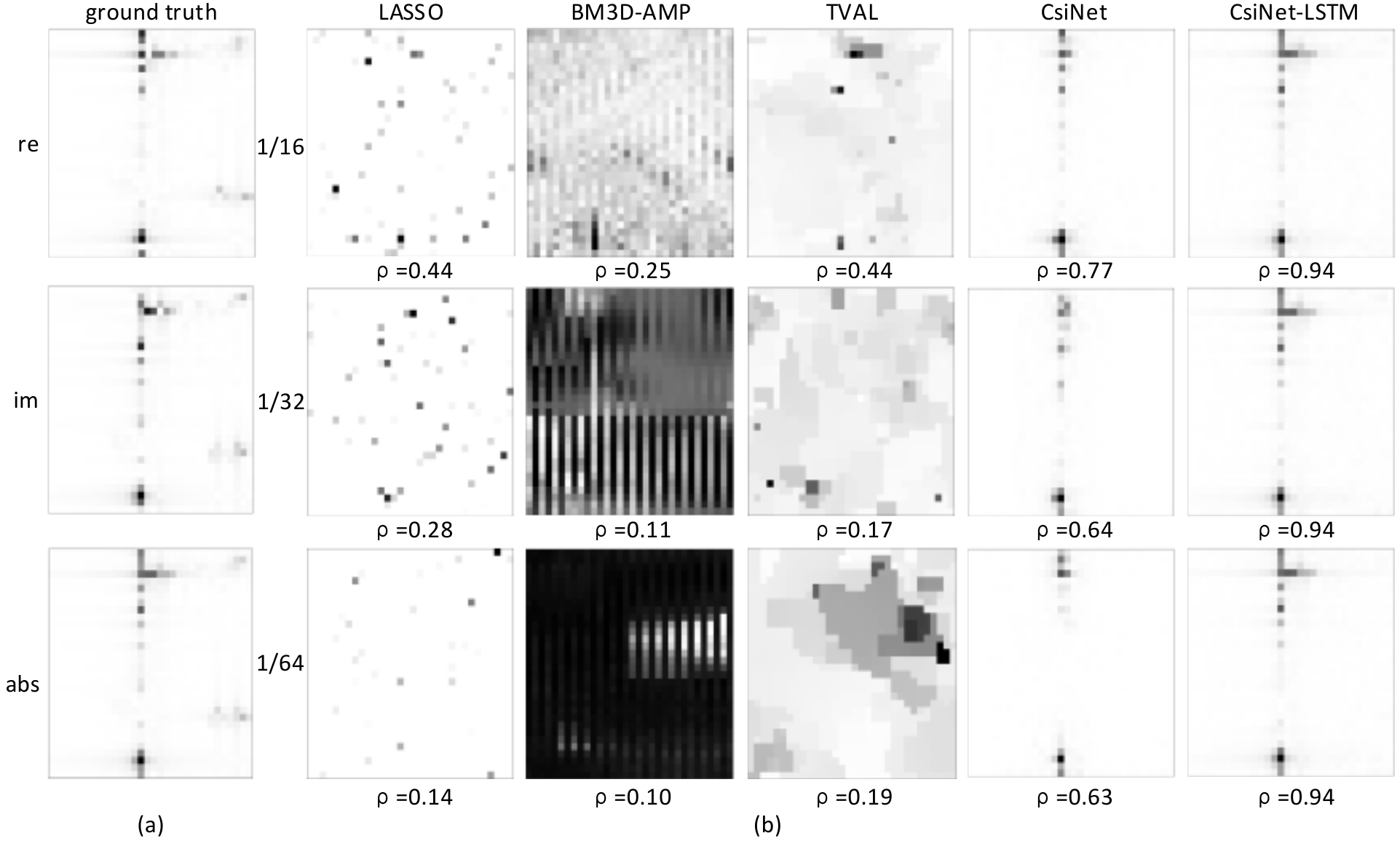}
			\vspace{-0.3cm}
			\caption{(a) Pseudo-gray plots of an original channel generated by COST 2100 model in outdoor scenario, showing real part, imagine part and absolute values, respectively. (b) Absolute values of reconstructed images, which are performed by different methods on the original channel given by (a) at different CRs.}
			\vspace{-0.5cm}
			\label{channel_plot}		
		\end{figure}

	The performance comparison of NMSE, $\rho$, and runtime are summarized in Table \ref{performance}. From the table, the DL-based CsiNet and CsiNet-LSTM considerably outperform all CS-based methods. Fig. \ref{channel_plot} gives a reconstruction result of the $5$th channel matrix of a certain channel group in outdoor scenario as an example, which represents the average performance at different CRs. Apparently, CsiNet and CsiNet-LSTM continue to offer adequate beamforming gain at low CRs, where CS-based methods fail to work. In particular, CsiNet-LSTM achieves the lowest NMSE at all CRs and is multiple times lower than CsiNet, especially when CR is low.
	
	Notebly, CsiNet-LSTM has the least performance loss as CR decreases, with only 8\% and 10\% for indoor and outdoor, respetively. The simulation results indicate that the remaining channel matrices  $\{{\bf H}_{t}''\}_{t=2}^{T}$ recovered from a low CR exhibit similar recovery quality and are better than the first channel matrix $ {\bf H}_{1}''$  from a high CR, which is $-14.74$ dB and $-8.35$ dB in average for the indoor and outdoor scenarios, respectively. This result is mainly attributed to the correlation of the channel matrices in time, which can be inherently retained by LSTMs. Moreover, since codewords are concatenated to offer more measurements before fed into the low-CR decoder, the remaining $T-1$ channel matrices achieve better recovery quality.
	\vspace{-0.35cm}
		\begin{table}[H]
			\caption{}
			\label{performance}
			\centering
			\begin{tabular}{m{0.1cm} | m{0.7cm}<{\centering} m{1cm}<{\centering} | m{0.8cm}<{\centering} m{0.8cm}<{\centering} m{0.7cm}<{\centering} m{0.75cm}<{\centering} m{0.75cm}<{\centering}}
				\hline
				\hline
				&  & CR & LASSO & BM3D-AMP & TVAL3 & CsiNet & CsiNet-LSTM \\
				\hline
				\multirow{10}{*}{\rotatebox{90}{Indoor}}& \multirow{3}{*}{NMSE} & 1/16 & -2.96 & 0.25 & -3.20 & -10.59 & \textbf{-23.06} \\
				& & 1/32 & -1.18 & 20.85 & -0.46 & -7.35 & \textbf{-22.33}\\
				& & 1/64 & -0.18 & 26.66 & 0.60 & -6.09 & \textbf{-21.24}\\
				\cline{2-8}
				& \multirow{3}{*}{$\rho$} & 1/16 & 0.72 & 0.29 & 0.73 & 0.95 & \textbf{0.99} \\
				& & 1/32 & 0.53 & 0.17 & 0.45 & 0.90 & \textbf{0.99}\\
				& & 1/64 & 0.30 & 0.16 & 0.24 & 0.87 & \textbf{0.99}\\
				\cline{2-8}
				& \multirow{3}{*}{runtime} & 1/16 & 0.2471 & 0.3454 & 0.3148 & \textbf{0.0001} & 0.0003 \\
				& & 1/32 & 0.2137 & 0.5556 & 0.3148 & \textbf{0.0001} & 0.0003\\
				& & 1/64 & 0.2479 & 0.6047 & 0.2860 & \textbf{0.0001} & 0.0003\\
				\cline{2-8}
				& NMSE$\downarrow$ & \scriptsize 1/16-\scriptsize 1/64 & 94\% & 105 & 1.19 & 42\% & \textbf{8\%} \\
				\hline
				\hline
				\multirow{10}{*}{\rotatebox{90}{Outdoor}}& \multirow{3}{*}{NMSE} & 1/16 & -1.09 & 0.40 & -0.53 & -3.60 & \textbf{-9.86} \\
				& & 1/32 & -0.27 & 18.99 & 0.42 & -2.14 & \textbf{-9.18}\\
				& & 1/64 & -0.06 & 24.42 & 0.74 & -1.65 & \textbf{-8.83}\\
				\cline{2-8}
				& \multirow{3}{*}{$\rho$} & 1/16 & 0.49 & 0.23 & 0.46 & 0.75 & \textbf{0.95} \\
				& & 1/32 & 0.32 & 0.16 & 0.28 & 0.63 & \textbf{0.94}\\
				& & 1/64 & 0.19 & 0.16 & 0.19 & 0.58 & \textbf{0.93}\\
				\cline{2-8}
				& \multirow{3}{*}{runtime} & 1/16 & 0.2122 & 0.4210 & 0.3145 & \textbf{0.0001} & 0.0003 \\
				& & 1/32 & 0.2409 & 0.6031 & 0.2985 & \textbf{0.0001} & 0.0003\\
				& & 1/64 & 0.0166 & 0.5980 & 0.2850 & \textbf{0.0001} & 0.0003\\
				\cline{2-8}
				& NMSE$\downarrow$ & \scriptsize 1/16-\scriptsize 1/64 & 94\% & 60 & 2.40 & 54\% & \textbf{10\%} \\
				\hline
				\hline			
			\end{tabular}
		\end{table}
		\vspace{-0.3cm}
	Furthermore, the DL-based methods benefit from GPU acceleration due to the feedforward and fast matrix vector multiplication nature, which perform approximately thousandfold faster than the CS-based methods. Compared with CsiNet, CsiNet-LSTM slightly loses time efficiency. However, its NMSE and $\rho$ are significantly improved. In addition, runtime is considerably shorter than the feedback interval $\delta t=0.04$ s, which makes real-time reconstruction possible.

	\vspace{-0.2cm}
	\section{Conclusion}
	\label{conclusion}
	In this article, we have proposed a real-time and end-to-end CSI feedback framework by extending the DL-based CsiNet with LSTM. CsiNet–LSTM achieves remarkable trade-off among CR, recovery quality, and complexity by utilizing the time correlation and structure properties of time-varying massive MIMO channels. We believe that this framework has the potential for practical deployment on real systems.
	\bibliographystyle{IEEEtran}	

\vspace{-0.3cm}
\begin{thebibliography}{10}
	\providecommand{\url}[1]{#1}
	\scriptsize
	\csname url@samestyle\endcsname
	\providecommand{\newblock}{\relax}
	\providecommand{\bibinfo}[2]{#2}
	\providecommand{\BIBentrySTDinterwordspacing}{\spaceskip=0pt\relax}
	\providecommand{\BIBentryALTinterwordstretchfactor}{4}
	\providecommand{\BIBentryALTinterwordspacing}{\spaceskip=\fontdimen2\font plus
		\BIBentryALTinterwordstretchfactor\fontdimen3\font minus
		\fontdimen4\font\relax}
	\providecommand{\BIBforeignlanguage}[2]{{%
			\expandafter\ifx\csname l@#1\endcsname\relax
			\typeout{** WARNING: IEEEtran.bst: No hyphenation pattern has been}%
			\typeout{** loaded for the language `#1'. Using the pattern for}%
			\typeout{** the default language instead.}%
			\else
			\language=\csname l@#1\endcsname
			\fi
			#2}}
	\providecommand{\BIBdecl}{\relax}
	\BIBdecl
	
	\bibitem{Love2008An}
	D.~J. Love, R.~W. Heath, V.~K.~N. Lau, D.~Gesbert, B.~D. Rao, and M.~Andrews,
	``An overview of limited feedback in wireless communication systems,''
	\emph{IEEE J. Sel. Areas Commum.}, vol.~26, no.~8, pp. 1341--1365, Oct. 2008.
	
	\bibitem{Kuo2012Compressive}
	P.~H. Kuo, H.~T. Kung, and P.~A. Ting, ``Compressive sensing based channel
	feedback protocols for spatially-correlated massive antenna arrays,'' in
	\emph{Proc. IEEE WCNC, Shanghai, China}, Apr. 2012, pp. 492--497.
	
	\bibitem{Rao2014Distributed}
	X.~Rao and V.~K. Lau, ``Distributed compressive {CSIT} estimation and feedback
	for {FDD} multi-user massive {MIMO} systems,'' \emph{IEEE Trans. Signal
		Process.}, no.~12, pp. 3261--3271, Jun. 2014.
	
	\bibitem{Wang2017Deep}
	T.~Wang, C.~K. Wen, H.~Wang, T.~Jiang, and S.~Jin, ``Deep learning for wireless
	physical layer: Opportunities and challenges,'' \emph{China Communications},
	vol.~14, no.~11, pp. 92--111, Nov. 2017.
	
	\bibitem{8052521}
	H.~Ye, G.~Y. Li, and B.~H. Juang, ``Power of deep learning for channel
	estimation and signal detection in {OFDM} systems,'' \emph{IEEE Wireless
		Communications Letters}, vol.~7, no.~1, pp. 114--117, Feb. 2018.
	
	\bibitem{He2018Deep}
	H.~He, C.~K. Wen, S.~Jin, and Y.~Li, ``Deep learning based channel estimation
	for beamspace mm{W}ave massive {MIMO} systems,'' \emph{IEEE Wireless Commun.
		Lett., to be published}, DOI 10.1109/LWC.2018.2832128.

	\bibitem{Wen2017Deep}
	C.~K. Wen, W.~T. Shih, and S.~Jin, ``Deep learning for massive {MIMO} {CSI}
	feedback,'' \emph{IEEE Wireless Commun. Lett., to be published}, DOI 10.1109/LWC.2018.2818160.
	
	\bibitem{Xu2018CSVideoNet}
	K.~Xu and F.~Ren, ``Csvideonet: A real-time end-to-end learning framework for
	high-frame-rate video compressive sensing,'' in \emph{Proc. IEEE WACV, NV,
		USA}, Mar. 2018, pp. 1680--1688.
	
	\bibitem{Wen2015Channel}
	C.~K. Wen, S.~Jin, K.~K. Wong, J.~C. Chen, and P.~Ting, ``Channel estimation
	for massive {MIMO} using gaussian-mixture {B}ayesian learning,'' \emph{IEEE
		Trans. Wireless Commun.}, vol.~14, no.~3, pp. 1356--1368, Mar. 2015.
	
	\bibitem{Liu2012The}
	L.~Liu, C.~Oestges, J.~Poutanen, and K.~Haneda, ``The {COST} 2100 {MIMO}
	channel model,'' \emph{IEEE Wireless Commun.}, vol.~19, no.~6, pp. 92--99,
	Dec. 2012.
	
	\bibitem{Daubechies}
	I.~Daubechies, M.~Defrise, and C.~D. Mol, ``An iterative thresholding algorithm
	for linear inverse problems with a sparsity constraint,'' \emph{Comm. Pure
		and Applied Math.}, vol.~75, pp. 1412--1457, Aug. 2004.
	
	\bibitem{Li2009User}
	C.~Li, W.~Yin, and Y.~Zhang, ``User’s guide for tval3: Tv minimization by
	augmented lagrangian and alternating direction algorithms,'' \emph{CAAM
		report}, vol.~20, pp. 46--47, 2009.
	
	\bibitem{Metzler2016From}
	C.~A. Metzler, A.~Maleki, and R.~G. Baraniuk, ``From denoising to compressed
	sensing,'' \emph{IEEE Trans. Inf. Theory}, vol.~62, no.~9, pp. 5117--5144,
	Sep. 2016.
	
\end{thebibliography}
		

\end{document}